\documentclass[twocolumn,           % Format : preprint, twocolumn
               showpacs,        10ppt,    % Pacs : showpacs, noshowpacs
               preprintnumbers,     % Preprint: preprintnumbers,
               aps,                 % Society: ...
               prd,          	    % Journal Style : pra, prb, prc, prd, pre,
               letterpaper,             % Size : a4paper, ...
               superscriptaddress,      % Affiliation (Title) : groupedaddress,
               nofootinbib,         % Footnote: footinbib, nofootinbib
               tightenlines,        % Remove additional spaces in a line
               floats,floatfix      % Floating pictures and tables
               ]{revtex4-1}
\usepackage{graphicx}  % needed for figures
\usepackage{dcolumn}   % needed for some tables
\usepackage{bm}        % for math
\usepackage{amsmath,amssymb}
\usepackage{float}
\usepackage{makecell}
	\usepackage{amsmath}
    \usepackage{mathrsfs}
	\usepackage{amssymb}
	\usepackage{graphicx} 
	\usepackage{makeidx}
	\usepackage{amsfonts}
	\usepackage[ansinew]{inputenc}
	\usepackage[usenames, dvipsnames]{pstricks}
	\usepackage{epsfig}
	\usepackage{pst-grad} % For gradients
	\usepackage{pst-plot} % For axes
	\usepackage[colorlinks, hyperindex]{hyperref}
	\usepackage{float}
	\usepackage{lipsum}
	\usepackage{makecell}
	\usepackage{ulem}
	\usepackage[caption=false]{subfig}
\usepackage{array}
\newcolumntype{P}[1]{>{\centering\arraybackslash}p{#1}}

\begin{document}

\title{ Dynamical systems analysis of phantom dark energy models}
\author{Nandan Roy} 
 \email{nandan@fisica.ugto.mx}
 \affiliation{%
Departamento de F\'isica, DCI, Campus Le\'on, Universidad de
Guanajuato, 37150, Le\'on, Guanajuato, M\'exico.}

\author{Nivedita Bhadra} 
 \email{nb12rs061@iiserkol.ac.in}
\affiliation{%
Department of Physical Sciences, IISER Kolkata, Nadia, West Bengal 741246, India}
%\begin{document}
%\title{{\Large\bf
% Stability analysis of Phantom dark energy models with different potentials}}
%
%\author{Nandan Roy} 
%\email{nandan@fisica.ugto.mx}
%\affiliation{%
%Departamento de F\'isica, DCI, Campus Le\'on, Universidad de
%Guanajuato, 37150, Le\'on, Guanajuato, M\'exico.}
%
%
%\author{Nivedita Bhadra} 
% \email{nandan@fisica.ugto.mx}
%\affiliation{%
%Department of Physical Sciences, IISER Kolkata, Nadia, West Bengal 741246, India}
%%

\begin{abstract} 

In this work, we study the dynamical systems analysis of phantom dark energy models considering five different potentials. From the analysis of these five potentials we have found a general parametrization of the scalar field potentials which is obeyed by many other potentials. Our investigation shows that there is only one fixed point which could be the beginning of the universe. However, future destiny has many possible options.  A detail numerical analysis of the system has been presented. The observed late time behaviour  in this analysis shows very well agreement with the recent observations.
\end{abstract}

\pacs{98.80.-k; 95.36.+x}

\keywords{cosmology, phantom dark energy, dynamical systems, phase space }

\maketitle
\section{Introduction}

 Several recent observations \cite{ade2016planck,perlmutter1999astrophys,riess1998observational,hinshaw2007three,tegmark2004cosmological,seljak2005cosmological} have confirmed the accelerated expansion of the universe. However, these observations do not offer any clear picture of the driver of this mysterious behaviour of the universe. The cosmological constant\citep{padmanabhan2003cosmological} is the most popular candidate which can successfully explain the late time acceleration but it suffers from its own problem which is known as the cosmological constant problem \cite{weinberg1989cosmological}.The alternative approaches \cite{copeland2006dynamics} to cosmological constant are generally classified into two different classes. In the first approach, the energy-momentum tensor is modified by introduction of an exotic matter with a negative pressure. These models are called  ``modified matter models". The second approach is called``modified gravity models" in which the gravity sector of the Einstein equation is modified. Among the modified matter models the quintessence\cite{cladwell,tsujikawa,copeland1998exponential} and phantom model \cite{tsujikawa, caldwell2002phantom, caldwell91phantom,ludwick2017viability} are very popular as the cosmological dynamics of these models have rich phenomenological behavior. 
In both quintessence and phantom model, a scalar field is minimally coupled to gravity and the potential supply the sufficient negative pressure to drive the accelerated expansion of the universe. The phantom scalar field has a negative kinetic term which is opposite to the quintessence scalar field model.
Present observations suggest that the equation of state of the dark energy is $\omega_{\phi} < -1$ \citep{ade2016planck}. In the conventional quintessence scalar field models, $\omega_{\phi} < -1$ is not achievable as these models are based on the canonical kinetic energy. A phantom field which has non-canonical kinetic term can give us a scenario of $\omega_{\phi} < -1$ in the evolution of the dark energy. There are some theoretical problems such as violation of some energy conditions \cite{ carroll2003can, cline2004phantom} which may occur due to the introduction of the phantom field but one can not deny the fact that it can very well fit the current observations \cite{ singh2003cosmological,sami2004phantom}. The literature is full of plenty of work on scalar field dark energy models with different potentials \cite{copeland1998exponential, zlatev1999quintessence, de2000cosmological, ng2001applications, corasaniti2003model, caldwell2005limits, linder2006paths, scherrer2008thawing, chiba2009slow}. None of these potentials enjoy any kind of preference from either theoretical or observational point of view. For a comprehensive list of potentials, we refer to \cite{sahni20045}.

  Several works have been done to study the scalar field dark energy models using dynamical systems analysis \cite{copeland1998exponential, roy2014quintessence, roy2014tracking, roy2015dynamical,dutta2016cosmological,Bhatia:2017wdh,Sola:2016hnq}. Dynamical systems analysis is a very useful method to study the qualitative behaviour of any non linear system. For a general discussion on the application of dynamical systems analysis in general relativity and cosmology, we refer to \cite{wainwright2005dynamical, coley2013dynamical,bahamonde2017dynamical}. The phase space behaviour of the phantom model with different potentials has been studied in \cite{guo2004attractor}  using the Hamilton-Jacobi formalism. This work considered a universe in which the phantom scalar field is the only component. In \cite{urena2005scalar}, Urena-Lopez studied the attractor  behavior of the phantom model with a positive exponential potential. A dynamical system analysis of a phantom model with a different scalar coupling functions and an exponential potential has been done in \cite{mahata2014dynamical}.
  
In this paper, we have done stability analysis of phantom models for five different potentials. By analyzing these five potentials we have found a general parametrization of the potential variable $\Gamma = (V \frac{d^2 V}{d \phi^2})/(\frac{dV}{d\phi})^2$. It is interesting to see that not only these five potentials many other potentials also follow this parametrization.

%In this paper, our aim is to perform a general stability analysis of the  phantom scalar field models. We have considered a general form of the $\Gamma = (V \frac{d^2 V}{d \phi^2})/(\frac{dV}{d\phi})^2$ of the potentials. The general form of the potential is written by observing the $\Gamma$ of five different type of potentials. Not only for these five potentials, this general form is valid for many other potentials 

%Lets write in more detail about the importance of these potentials . To the best of my understanding, we have performed for these five only as an example. Each one shows the same results or predicts similar late time behaviour. We have to probably emphasize on that. ???}). 

Using this general parametrization we have tried to make the analysis as much general as possible . We have found out fixed points and corresponding eigenvalues of the system considering the general form. But it seems difficult to find out the stability conditions for some fixed points as the mathematical expression of the eigenvalues are very long and complicated. So the stability analysis has been done separately for all the five potentials which we have considered to write the general form.  A numerical investigation of the system has been done considering the general form.

A summary of the contents of this paper is as follows: Section II is a brief introduction to the mathematical background of the phantom model. Section III is the stability analysis of the system with different type of potentials. Section IV is a numerical investigation of the system and discussion about the results obtained. 

\section{Mathematical background}
We consider a universe which is homogeneous, isotropic and spatially flat. This type of universe is mathematically represented by flat FRW metric
\begin{align}
ds^2=dt^2-a(t)^2(dr^2+r^2d\Omega^2).
\end{align}

 We also consider in this universe, the matter sector is dominated by a barotropic fluid with an equation of state $p_m=(\gamma-1) \rho_m$, where $\gamma$ is the equation of state parameter, $\rho_m$ is the energy density of the perfect fluid and $p_m$ is the corresponding pressure.  In addition to the  matter, the universe is also filled by a phantom scalar field which is minimally coupled to gravity. The action of this minimally coupled phantom scalar field can be written as 

\begin{equation}
\mathcal{S} = \int d^4 x \sqrt{- g} (\frac{1}{2} R - \frac{1}{2} g^{\mu \nu} \partial_{\mu} \phi  \partial^{\nu} \phi +V(\phi)),
\end{equation}

where $\phi$ is the phantom field and $V(\phi)$ is the scalar potential. By varying the action with respect to the metric one can get the Friedmann equations as
\begin{align}
&H^2=\frac{8\pi G}{3}(\rho_m-\frac{1}{2}\dot{\phi}^2+V(\phi)), \label{Fri1}\\
&\dot{H}=-\frac{8\pi G}{2}( \gamma \rho_m - \dot{\phi}^2) \label{Fri2},
\end{align}
 where, $H=\frac{\dot{a}}{a}$ is the Hubble parameter, $a(t)$ is the scale factor of the universe and dot means the differentiation with respect to time. The conservation equation for the fluid is,
\begin{align}
\dot{\rho}_m=-3 \gamma H \rho_m. \label{continuity}
\end{align}
The wave equation is given by,
\begin{align}
\ddot{\phi}+3H\dot{\phi}=\frac{dV}{d\phi}. \label{wave}
\end{align}
Our aim is to study the phase space behaviour of this model and for that, the system has to be written as a set of autonomous equations. We define following dimensionless variables as,
\begin{align}
x=\frac{k \phi'}{\sqrt{6}},\; y^2=\frac{k^2 V}{3H^2},
\end{align}
where, the prime represents the differentiation with respect to $N=\ln (\frac{a}{a_0})$ and $k^2=8\pi G$. The present value of the scale factor, $a_0$, is chosen as unity. The energy density and the effective pressure due to the phantom field can be written as,
\begin{align}
&\rho_\phi= - \frac{1}{2}\dot{\phi}^2+V(\phi),\\
&p_\phi=- \frac{1}{2}\dot{\phi}^2-V(\phi).
\end{align}
We also assume that $p_{\phi}$ and $\rho_{\phi}$ obey the barotropic relation, $p_\phi=(\gamma_{\phi}-1)\rho_\phi$.
Thus the equation of state parameter $\gamma_\phi$ for the scalar field in terms of dimensionless variables can be written as,
\begin{align}
\gamma_\phi= \frac{\rho_\phi+p_\phi}{\rho_\phi}=  \frac{- \dot{\phi}^2}{- \frac{\dot{\phi}^2}{2}+V}=\frac{- 2x^2}{- x^2 + y^2}.
\end{align}

The density parameter $\Omega_\phi$ for the scalar field is given by,
\begin{align}
\Omega_\phi=\frac{k^2 \rho_\phi}{3H^2}= - x^2+y^2,
\end{align}
which is restricted by the Friedmann constrain equation, (\ref{Fri1}) as ,
\begin{align}
\Omega_m+\Omega_\phi=1,
\end{align}
$\Omega_m = \frac{\kappa^2 \rho_{m}}{3 H^2}$, is baryonic energy density parameter. \par
As we are interested in the late time behaviour of the universe so from now onwards we consider $\gamma = 1$, which describe a matter dominated universe. Then the deceleration parameter $(q=-\frac{a\ddot{a}}{a^2}=-\frac{\dot{H}+H^2}{H^2})$ can be expressed as,
\begin{align}
q=\frac{3 }{2}(1+x^2-y^2)- 3x^2-1
.\end{align}
Using equation (\ref{Fri1}, \ref{Fri2}, \ref{wave}), the system of equations can be rewritten  as an autonomous system in terms of these new variables,
\begin{align}
\label{eq:x} 
&x'=-3x-\lambda\sqrt{\frac{3}{2}}y^2+\frac{3}{2}x[(1-x^2-y^2)],\\
\label{eq:y} 
&y'=-\lambda\sqrt{\frac{3}{2}}xy+\frac{3}{2}y[ (1-x^2-y^2)],\\
\label{eq:lam} 
&\lambda'=-\sqrt{6}\lambda^2(\Gamma-1)x = - \sqrt{6} x f,
\end{align}
where, $\lambda=-\frac{1}{kV}\frac{dV}{d\phi},  \Gamma=V\frac{d^2 V}{d\phi^2}/(\frac{dV}{d\phi})^2$ and $f= \lambda^2 (\Gamma -1)$.

 To close the system of equations (\ref{eq:x}, \ref{eq:y}, \ref{eq:lam}) one needs to know the particular form of the $f$ which depends on particular choice of potentials. The choice of the potentials remains arbitrary until there is a selection of a potential by the fundamental physics or by cosmological observation. So there is no particular form of the potential which is a natural choice. In this work, as for examples, we have considered five different type of potentials for which $f$ can be written as a function of $\lambda$. Corresponding $f$  of these potentials are listed in Table I. From Table I, one can see that the form of  $f$ has a general structure which is of the form $f(\lambda)=\alpha_1 \lambda^2+\alpha_2 \lambda+\alpha_3$. Not only these five potentials there is a list of potentials which follow this general form of the $f$. For a comprehensive list of these potentials, we refer to \cite{fang2009exact, bahamonde2017dynamical}. These potentials are very often used in quintessence scalar field models and has a rich phenomenological behaviour (see references in Table I).

\begin{table*}[htb]
\caption{List of potentials and corresponding $f$.}
\begin{ruledtabular}
\centering
\begin{tabular}{ |P{2cm}||P{2cm}|P{2cm}|P{2cm}| P{2cm}| P{2cm}}
 \multicolumn{6}{|c|}{List of potential} \\
 References& Potential $V(\phi)$& $f$ &$\alpha_1$&$\alpha_2$&$\alpha_3$\\
 \hline
A \cite{frieman1995cosmology, linde1983chaotic} & $V_0\phi^n$  &$-\frac{\lambda^2}{n}$& $-\frac{1}{n}$    &0&0\\
 B \cite{zlatev1999quintessence} & $V_0 e^{-k\phi}+V_1$&   $-\lambda^2+k\lambda$  & $-1$   &$k$&$0$\\
C \cite{sahni2000new}&  $\cosh(\xi \phi)-1$ &$-\frac{1}{2}\lambda^2+\frac{1}{2}\xi^2$ & $-\frac{1}{2}$&$0$&$\frac{1}{2}\xi^2$\\
D \cite{sahni2000case,urena2000new}& $V_0 \sinh^{-\alpha}(\beta \phi)$ &$\frac{\lambda^2}{\alpha}-\alpha \beta^2$& $\frac{1}{\alpha}$&  $0$&$-\alpha \beta^2$\\
E \cite{urena2016new} & $2 M^2 \cos (\frac{\phi}{2 l})^2$&  $-\frac{1}{2}\lambda^2-\frac{1}{2l^2}$  & $-\frac{1}{2}$&$0$&$-\frac{1}{2l^2}$\\
 \end{tabular}
\end{ruledtabular}
\end{table*}

\begin{table*}[]
\caption{List of the fixed points and the cosmological parameters. \label{tab:1}}
\begin{ruledtabular}
\begin{tabular}{|P{1cm}|P{1cm}|P{2cm}|P{2cm}|P{2cm}|P{1cm}|P{4cm}|}
 \multicolumn{7}{|c|}{List of fixed points} \\
 \hline
 Fixed points & $x$&$y$&$\lambda$& $q$ &    $\Omega_{\phi}$& Eigenvalues\\
 \hline\hline
 $ a$ & $0 $ & $0 $ & Undetermined & $ q = \frac{1}{2} $ & 0&$0,\frac{3}{2},- \frac{3}{2}$\\
$b$ &$0$& $+1$ & $0$ & $ q =  -1$ & 1& $-3, \frac{1}{2} (-3 \pm \sqrt{12 \alpha _3+9} )$\\
$c$ &$0$& $-1$ & $0$ & $ q =  -1$ & 1 &$-3, \frac{1}{2} (-3 \pm \sqrt{12 \alpha _3+9} )$\\
$d$ &$-\frac{\lambda_{+}}{\sqrt{6}}$& $\sqrt{1+\frac{\lambda_{+}^2}{6}}$ & $\lambda_{+}$ & $ q =  - \frac{\lambda_{+}^2}{2} -1$ & 1& $- 3 - \lambda_+^2, m_d, n_d$\\
$e$ &$-\frac{\lambda_{-}}{\sqrt{6}}$& $\sqrt{1+\frac{\lambda_{-}^2}{6}}$ & $\lambda_{-}$ & $ q =  - \frac{\lambda_{-}^2}{2} -1$ & 1&$- 3 - \lambda_-^2, m_e, n_e$\\
$f$ &$-\frac{\lambda_{+}}{\sqrt{6}}$& $-\sqrt{1+\frac{\lambda_{+} ^2}{6}}$ & $\lambda_{+}$ & $ q =  - \frac{\lambda_{+}^2}{2} -1$ & 1&$- 3 - \lambda_+^2, m_f, n_f$\\
$g$ &$-\frac{\lambda_{-}}{\sqrt{6}}$& $-\sqrt{1+\frac{\lambda_{-}^2}{6}}$ & $\lambda_{-}$ & $ q =  - \frac{\lambda_{-}^2}{2} -1$ & 1 &$- 3 - \lambda_-^2, m_g, n_g$\\
\end{tabular}
 \end{ruledtabular} 
\end{table*}

To find the stability of the system we need to find out the fixed points of the system. The fixed points of the system are the simultaneous solutions of the equations $x^\prime = 0, y^\prime = 0, \lambda^\prime = 0$.   The list of the fixed points, their corresponding eigenvalues and corresponding cosmological parameters are given in the Table 2 considering the general form of $f(\lambda)=\alpha_1 \lambda^2+\alpha_2 \lambda+\alpha_3$. 

Here, $\lambda_{\pm} = \frac{1}{2 \alpha_1} (-\alpha_2 \pm \sqrt{\alpha_2 ^2 - 4 \alpha_1 \alpha_3})$ are the two solutions of the quadratic equation $f(\lambda)=    \alpha_1 \lambda^2+\alpha_2 \lambda+ \alpha_3 = 0$.  Fixed point $a$ belongs to a special class of nonhyperbolic fixed points which is called normally hyperbolic fixed point. A normally hyperbolic fixed point is a set of fixed points which are nonisolated in nature and has one zero eigenvalue at each fixed point. The stability of a normally hyperbolic fixed point can be found out from the sign of the remaining eigenvalues of the fixed point. Negative sign corresponds to a stable fixed point,  positive sign corresponds to an unstable fixed point and a mixture of both corresponds to a saddle fixed point.  One can see that the fixed point $a$ is always saddle and it does not depend on any particular choice of the potentials. This fixed point is matter dominated and decelerated. Fixed point $b$ and $c$ are always dark energy dominated and accelerated.  For $\alpha_3 \neq 0$, they are always saddle in nature. For $\alpha_3 = 0$ these fixed points becomes non-hyperbolic as the eigenvalues are $(-3,-3,0)$ and we need a numerical approach to find the stability. Two eigenvalues of the fixed points $d$ to $g$ are denoted by $m$ and $n$ as they are very big and complected (please see appendix). It is very difficult to draw  any conclusion about the stability of these fixed points without considering any particular form of the potential. In the next section, we consider the same potentials from the Table 1 as for example to study the stability of these fixed points.

\section{Stability analysis with  different potentials.}

\subsection{ $V(\phi)=V_0\phi^n$.} 

For this potential $f(\lambda) = - \frac{\lambda^2}{n}$. Though all the fixed points are allowed for this potential but $b,d,e$ (0,1,0)  and $c,f,g$ (0,-1,0 ) becomes indistinguishable. Basically, there are three fixed points $a,b$ and $c$. The eigenvalues of the fixed points for this potential is given in Table III. All of these fixed points are nonhyperbolic in nature as out of the three eigenvalues one eigenvalue is zero. The stability of the fixed point $a$ has been already discussed in the previous section. For other two fixed points we cannot use the same method as these fixed points are not normally hyperbolic and also we cannot implement standard linear stability analysis for these nonhyperbolic fixed points. This difficulty can be overcome by numerical investigation of the stability. In order to do that generally the system is perturbed from each fixed point and allowed to evolved numerically. If the system comes back to the fixed point, the fixed point is stable; otherwise unstable. 

To check the stability of the fixed points $b$ and $c$, the evolution of the perturbations around these fixed points are plotted. It is very difficult to draw any physical conclusion from the 3D plot as it is very obscure, so for the sake of simplicity, we have shown the evolution of the variables $x,y,\lambda$ individually in Fig. 1.  From Fig 1 one can see that the perturbations around fixed point $b$ indeed comes back to the fixed point $b$. In Fig 2 the perturbations around the fixed point $c$ also show similar behavior. These plots are for $ n=1$. For $n=2$ the qualitative behaviour of the plots are similar. Hence we conclude that the fixed point $b$ and $c$ are stable fixed points.  Fixed point $a$ is a matter dominated point and always decelerated. Fixed point $b$ and $c$ are scalar field dominated and the universe at these points always expands with a constant acceleration $q = -1$. Due to the saddle nature of the fixed point $a$, it could be the beginning of the universe which means that the universe started from a matter dominated decelerated state. Fixed point $b$ and $c$ are late time attractors and these points may be the ultimate fate of the universe which indicates a universe
in future completely dominated by
the dark energy and ever accelerating..

\begin{figure*}[]
\begin{center}

\subfloat[Projection of perturbations along $x$ axis.]{%
  \includegraphics[clip,width=0.6\columnwidth]{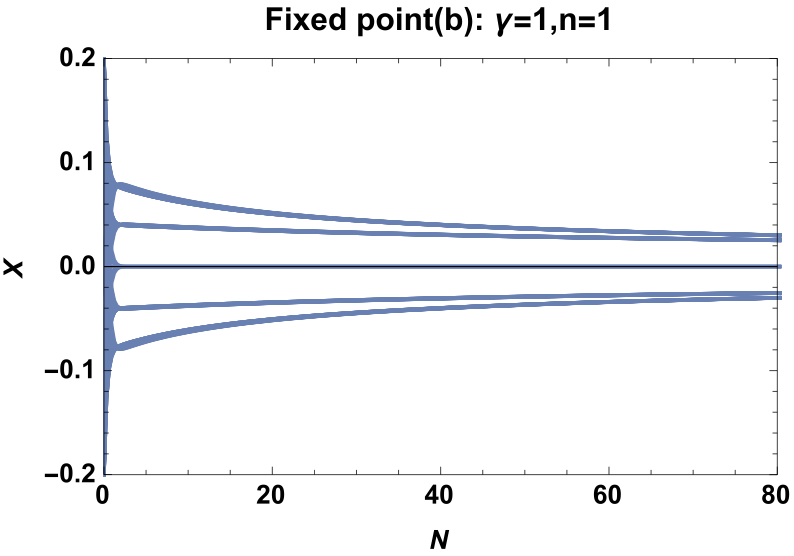}%
}
\subfloat[Projection of perturbations along $y$ axis.]{%
  \includegraphics[clip,width=0.6\columnwidth]{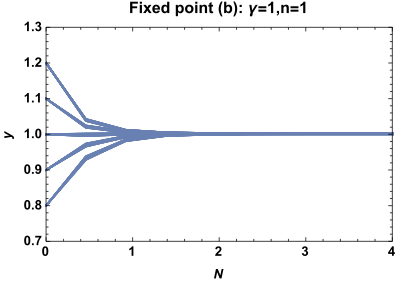}%
}
\subfloat[Projection of perturbations along $\lambda$ axis.]{%
  \includegraphics[clip,width=0.6\columnwidth]{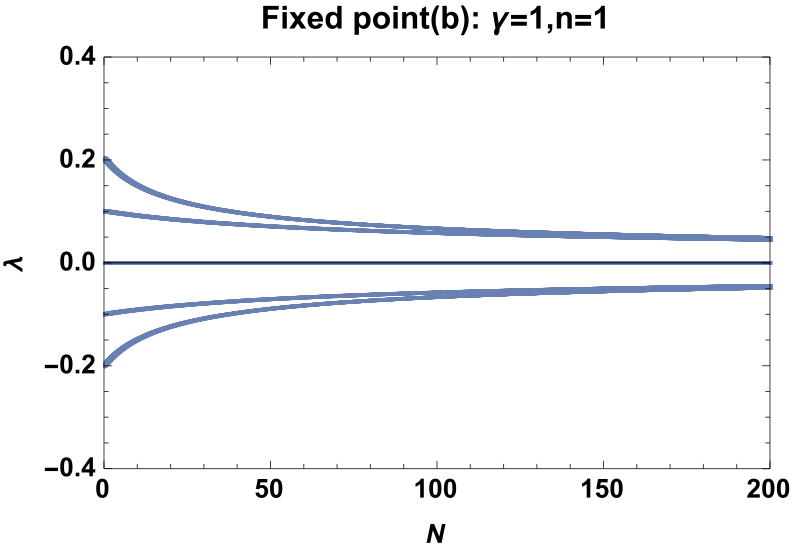}%
}

\caption{Projection of perturbations along $x,y , \lambda$ axis for potential $V(\phi)=V_0\phi^n$ with $\gamma =1, n =1$ around the fixed point b.}

\end{center}

\end{figure*}

\begin{figure*}[]
\begin{center}

\subfloat[Projection of perturbations along $x$ axis.]{%
  \includegraphics[clip,width=0.6\columnwidth]{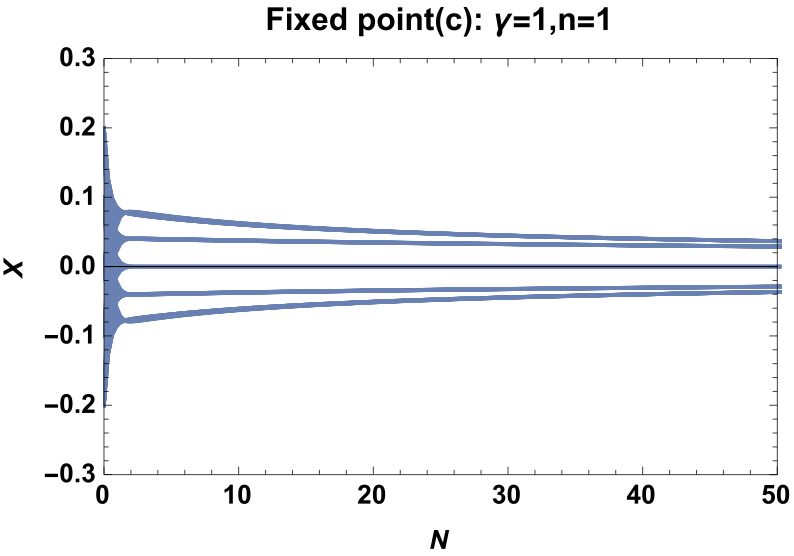}%
}
\subfloat[Projection of perturbations along $y$ axis.]{%
  \includegraphics[clip,width=0.6\columnwidth]{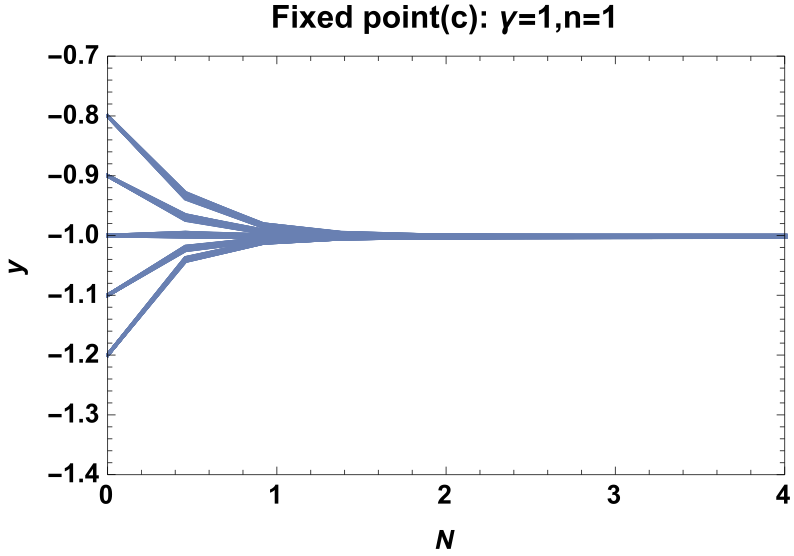}%
}
\subfloat[Projection of perturbations along $\lambda$ axis.]{%
  \includegraphics[clip,width=0.6\columnwidth]{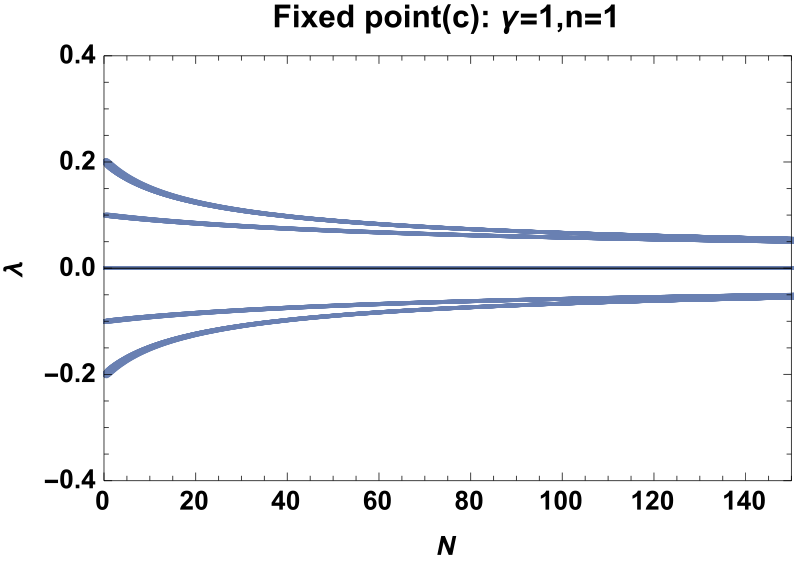}%
}

\caption{Projection of perturbations along $x,y , \lambda$ axis for potential $V(\phi)=V_0\phi^n$ with $\gamma =1, n =1$ around the fixed point c.}

\end{center}

\end{figure*}

\begin{table}[htp!]
\caption{Fixed points and the corresponding eigenvalues of the potential $A:  V(\phi) = V_0\phi^n$}
\begin{ruledtabular}
\centering
\begin{tabular}{|p{3cm}|p{3cm}|}
 \multicolumn{2}{|c|}{Fixed points and Eigenvalues} \\
 \hline\hline
Fixed points&Eigenvalues\\ 
\hline
$a$&$0,\frac{3}{2},-\frac{3}{2}$\\
$b,d,e$&$-3,0,-3$\\
$c,f,g$&$-3,0,-3$\\
\end{tabular}
\end{ruledtabular}
\end{table}

\subsection{$V(\phi)=V_0 e^{-k\phi}+V_1$} 

 From the expression of $f$, we get
 \begin{eqnarray}
 \lambda_{\pm}&=&-\frac{k\pm \sqrt{k^2}}{2},\nonumber\\
 \lambda_{+}&=&-k,\nonumber\\
 \lambda_{-}&=&0.
  \end{eqnarray}

The fixed points and the eigenvalues are given in Table IV. Fixed point $a$ is normally hyperbolic and saddle in nature. $b,c$ are nonhyperbolic and need to be dealt with numerically.  Other fixed points are hyperbolic fixed points. One can see from the Fig 3 and Fig 4 that some perturbations do not come back to the fixed point $b$ and $c$. Apparently, it seems that these fixed points are saddle in nature. For different values of $k$  nature of these fixed points remains same. Fixed point $d$ and $f$ are always attractor and dominated by the dark energy. For this potential there are three possible beginning of the universe, fixed point $a,b$ and $c$. Fixed points $b$ and $c$ are saddle so a hateroclinic solution may start from these fixed points but these fixed points are accelerating and dark energy dominated. A dark energy dominated accelerated beginning of the universe is not supported by the observations. So from the observational point of view fixed point $a$ remains as the best choice to be the beginning of the universe. Like the power law potentials this potentials also have similar behaviour which is a decelerated matter dominated beginning and a dark energy dominated ever-accelerating future.

\begin{figure*}[]
\begin{center}

\subfloat[Projection of perturbations along $x$ axis.]{%
  \includegraphics[clip,width=0.6\columnwidth]{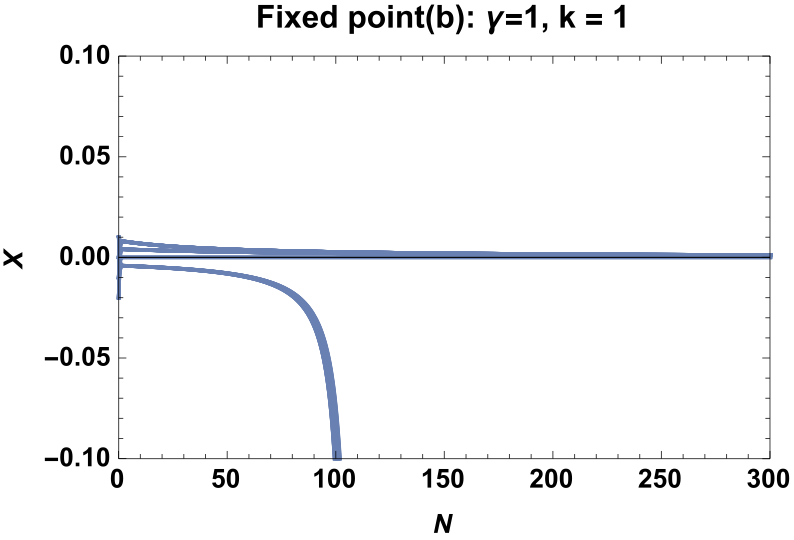}%
}
\subfloat[Projection of perturbations along $y$ axis.]{%
  \includegraphics[clip,width=0.6\columnwidth]{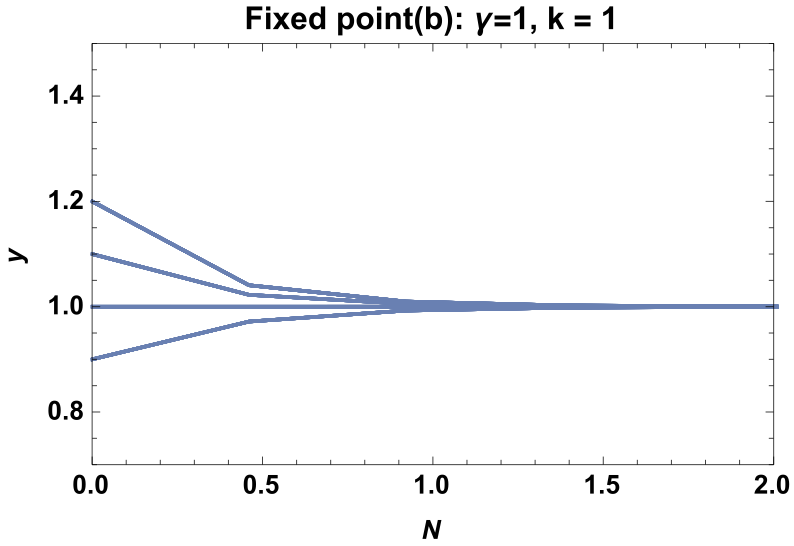}%
}
\subfloat[Projection of perturbations along $\lambda$ axis.]{%
  \includegraphics[clip,width=0.6\columnwidth]{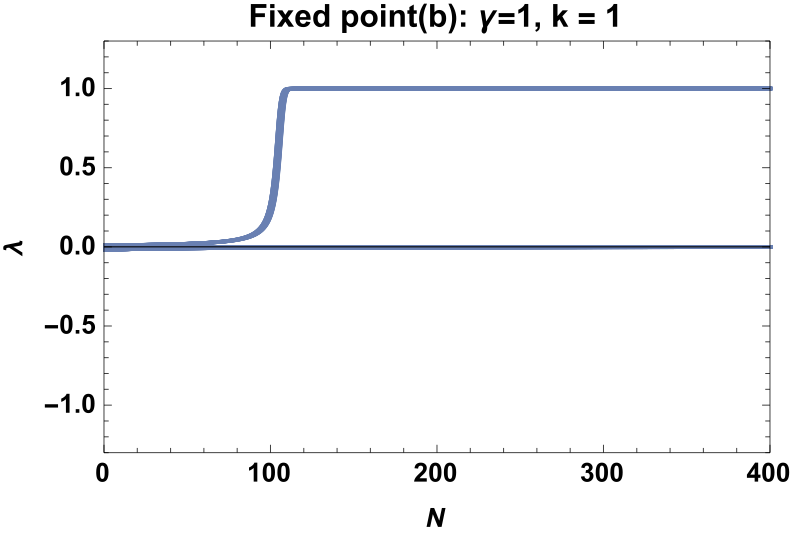}%
}

\caption{Projection of perturbations along $x,y , \lambda$ axis for potential $V(\phi)=V_0 e^{-k\phi}+V_1$ with $\gamma =1, k =1$ around the fixed point b.}

\end{center}

\end{figure*}

\begin{figure*}[]
\begin{center}

\subfloat[Projection of perturbations along $x$ axis.]{%
  \includegraphics[clip,width=0.6\columnwidth]{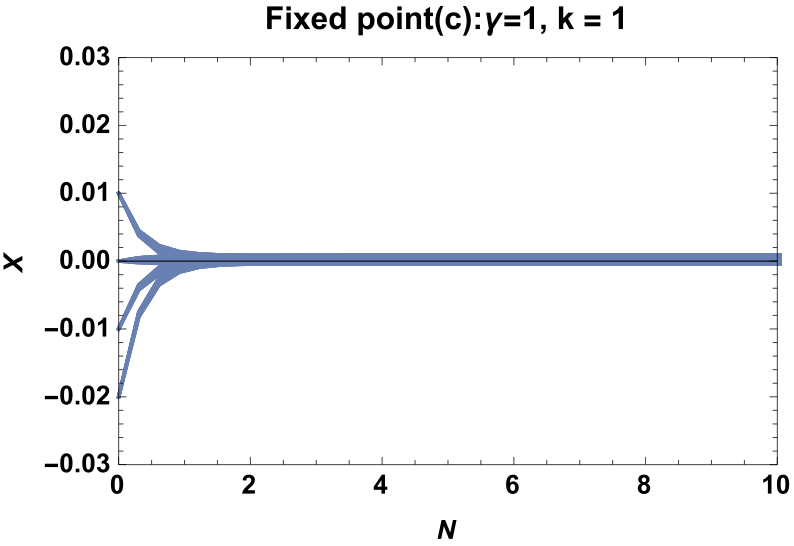}%
}
\subfloat[Projection of perturbations along $y$ axis.]{%
  \includegraphics[clip,width=0.6\columnwidth]{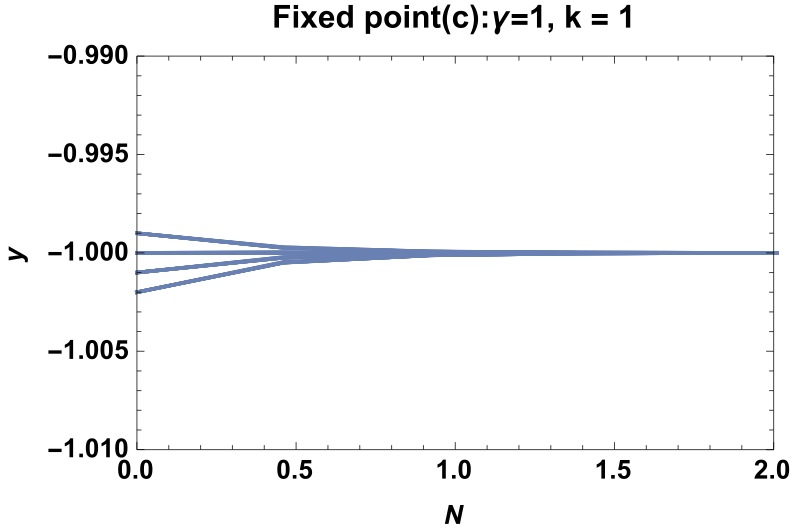}%
}
\subfloat[Projection of perturbations along $\lambda$ axis.]{%
  \includegraphics[clip,width=0.6\columnwidth]{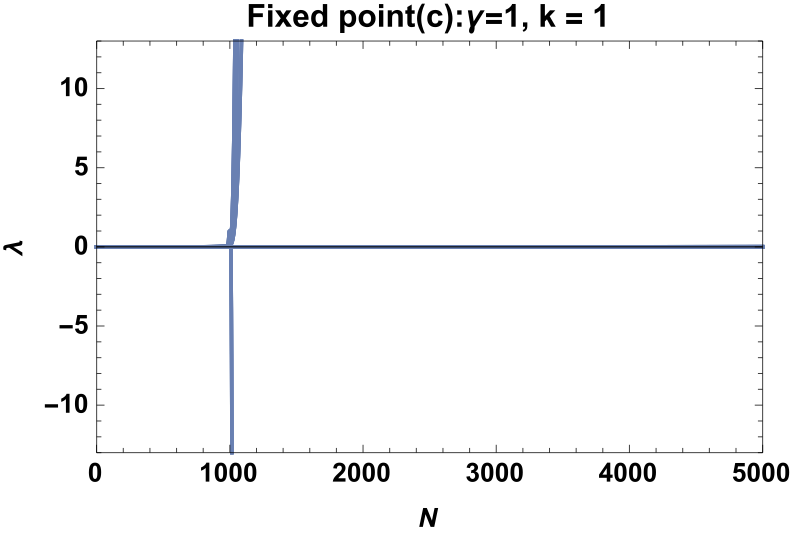}%
}

\caption{Projection of perturbations along $x,y , \lambda$ axis for potential $V(\phi)=V_0 e^{-k\phi}+V_1$ with $\gamma =1, k =1$ around the fixed point c.}

\end{center}

\end{figure*}

 \begin{center}
 \begin{table}[]
 \caption{Fixed points and the corresponding eigenvalues of the potential  B :$V(\phi)=V_0 e^{-k\phi}+V_1$}
 \begin{ruledtabular}
\begin{tabular}{|p{1cm}|p{7.5cm}|}
 \multicolumn{2}{|c|}{Fixed points and Eigenvalues} \\
 \hline
Fixed points&Eigenvalues\\ 
\hline
$a$&$0,\frac{3}{2},-\frac{3}{2}$\\
$b,e$&$-3,0,-3$\\
$c,g$&$-3,0,-3$\\
$d$& $-k^2,-\frac{1}{2}(6+k^2),-(k^2+3$)\\
$f$&$ -k^2,-\frac{1}{2}(6+k^2),-(k^2+3$) \\
\end{tabular}
\end{ruledtabular}
\end{table}
\end{center}
\subsection{$V(\phi)= \cosh(\xi \phi)-1$.  .} 
 
  For this potential $\lambda_{+}=\xi$ and $\lambda_{-}=-\xi$. All the fixed points are allowed and the eigenvalues are listed in the Table V.  Fixed point $a$ is saddle. Fixed point $b$ and $c$ have same eigenvalues and these fixed points are also always saddle. Fixed points $d,f,e,$ and $g$ are always stable. As the fixed point $d,f,e,$ and $g$ are attractor so the universe may be attracted towards these fixed points which are dark energy dominated and ever accelerating.
  
\begin{center}
 \begin{table}[]
 \caption{Fixed points and the corresponding eigenvalues of the potential C: $V(\phi)= \cosh(\xi \phi)-1$.}
 \begin{ruledtabular}
\begin{tabular}{|p{1cm}|p{8cm}}
% \multicolumn{2}{|c|}{Fixed points and Eigenvalues} \\
Fixed points&Eigenvalues\\ 
\hline
$a$&$0,\frac{3}{2},-\frac{3}{2}$\\
$b$ &$-3,-\frac{1}{2}\Big(3+3\sqrt{3(3+2\xi^2)}\Big),-\frac{1}{2}\Big(3-3\sqrt{3(3+2\xi^2)}\Big)$\\
$c$&$-3,-\frac{1}{2}\Big(3+3\sqrt{3(3+2\xi^2)}\Big),-\frac{1}{2}\Big(3-3\sqrt{3(3+2\xi^2)}\Big)$\\
$d,f$&$-\xi^2,-\frac{1}{2}(6+\xi^2),-(3+\xi^2)$\\
$e,g$&$-\xi^2,-\frac{1}{2}(6+\xi^2),-(3+\xi^2)$\\
\end{tabular}
\end{ruledtabular}
\end{table}
\end{center}
\subsection{$V(\phi)=V_0 \sinh^{-\alpha}(\beta \phi),$}. Here, $\lambda_{+}=\alpha \beta,\lambda_{-}=-\alpha \beta$.
All the fixed points exist for this potential. The corresponding eigenvalues of the fixed points are given in Table VI.  The fixed points b and c are stable for $\frac{2}{3} < \alpha \beta^2 $ and these fixed points are spiral attractor for $\alpha \beta^2 > \frac{3}{4}$. Fixed point $d,e,f,g $ are stable only if $\alpha <0$. The stability analysis of this potential also indicates a decelerated matter dominated beginning of the universe and an accelerated dark energy dominated future.

\begin{center}
 \begin{table}[]
 \squeezetable
 \caption{Fixed points and the corresponding eigenvalues of the potential  D: $V(\phi)=V_0 \sinh^{-\alpha}(\beta \phi)$}
 \begin{ruledtabular}
\begin{tabular}{|p{1cm}|p{8cm}}
% \multicolumn{2}{|c|}{Fixed points and Eigenvalues} \\
Fixed points&Eigenvalues\\ 
\hline
$a$&$0,\frac{3}{2},-\frac{3}{2}$\\
$b$&$-3,-\frac{1}{2}\Big(3+\sqrt{3(3-4\alpha\beta^2)} \Big),-\frac{1}{2}\Big(3-\sqrt{3(3-4\alpha\beta^2)} \Big)$\\
$c$&$a,b$\\
$d,e,f,g$& $2\alpha\beta^2,-\frac{1}{2}(6+\alpha^2\beta^2),-(\alpha^2\beta^2+3)$\\
\end{tabular}
\end{ruledtabular}
\end{table}
\end{center}
\subsection{$V(\phi)=2M^2\cos(\frac{\phi}{2l})^2$} 

Only fixed point a,b and c exist for this potential, where the saddle nature of the fixed point a remains same. Other two fixed points have same eigenvalues and they are always saddle, see Table VII. The cosmological dynamics, in this case, is very simple. Fixed point a could be the beginning of the universe and there is no late time attractor for this potential. 

\begin{center}
\begin{table}[H]
\caption{Fixed points and the corresponding eigenvalues of the potential  E:  $V(\phi)=2M^2\cos(\frac{\phi}{2l})^2$}
\begin{ruledtabular}
\begin{tabular}{|p{1cm}|p{8cm}}
 %\multicolumn{2}{|c|}{Fixed points and Eigenvalues} \\
 \hline
Fixed points&Eigenvalues\\ 
\hline 
$a$&$0,\frac{3}{2},-\frac{3}{2}$\\
$b$&$-\frac{1}{2}(3+\sqrt{3(3+\frac{2}{l^2})})$,$-\frac{1}{2}(3-\sqrt{3(3+\frac{2}{l^2})})$,$-3$\\
$c$&$-\frac{1}{2}(3+\sqrt{3(3+\frac{2}{l^2})})$,$-\frac{1}{2}(3-\sqrt{3(3+\frac{2}{l^2})}), - 3 $\\
\end{tabular}
\end{ruledtabular}
\end{table}
\end{center}

\section{Numerical Investigation}

In this section, we discuss the numerical integration of the system. One can see from the Table I that fixed point $a$ is always saddle,  matter dominated and decelerated. This nature of the fixed point $a$ is same for all kind of potentials which follows the general parametrization of $f$ as the stability condition of the fixed point does not depend on the particular form of the potentials.

 As it is matter dominated and decelerated  from the observational point of view fixed point $a$ is the best choice as the beginning of the universe. In a phase space, a heteroclinic solution joins an unstable or saddle fixed points to a stable fixed point. There will be no evolution of the system if we start exactly from the fixed point. So in this numerical investigation, we have allowed the system to evolve from the neighborhood of the fixed point $a$. The numerical integration is very stiff and we found it to be integrable for a very small parameter range $[\alpha_1, -1 : 0.6], [\alpha_2, 0:4], [\alpha_3, -2:0] $.  It deserves mention that this range of parameters includes all the possibilities to get back the potentials in the Table I. So this numerical analysis represents a general analysis of all the potentials listed in Table I. From Fig 5, one can see that the solutions around the fixed point $a$ are attracted towards the fixed point $b$. Fig 6, is the plot of the density parameters and it can be seen that the universe is now dark energy dominated and $\Omega_{\phi} \simeq 0.67$. The plot of the deceleration parameter $q$ in the Fig. 7 also shows a that the universe has smoothly entered into an accelerated expansion phase from a decelerated expansion phase around $z \simeq 0.55 $ which is in good agreement with the observations. Fig8 shows that the equation of state of the scalar field is slightly lower than $-1$ and which is much supported by observations (\cite{Ade:2015xua}).  These solutions are originated from the fixed point $a$ and it is interesting to see that these solutions can describe the accelerated expansion of the universe and satisfy the current observations for a general class of potentials which obey $f(\lambda)=\alpha_1 \lambda^2+\alpha_2 \lambda+\alpha_3$.

\begin{figure}[h]
\includegraphics[scale=0.6]{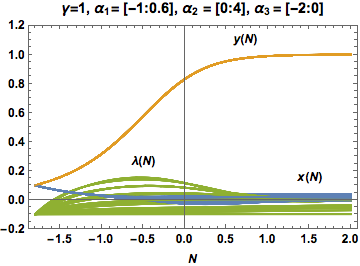} 
\caption{Plot of $x(N), y(N)$ and $\lambda(N)$ against $N = \ln a$ from the fixed point a .}

\end{figure}

\begin{figure}[h]
\centering
\includegraphics[scale=0.7]{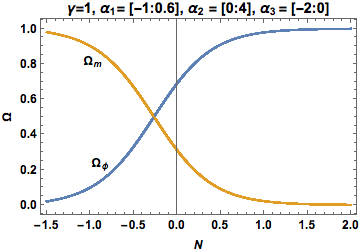} 
\caption{Plot of density parameters $\Omega_m$ and $\Omega_{\phi}$ against $N = \ln a$ from the fixed point a.}

\end{figure}

\begin{figure}[H]
\centering
\includegraphics[scale=0.7]{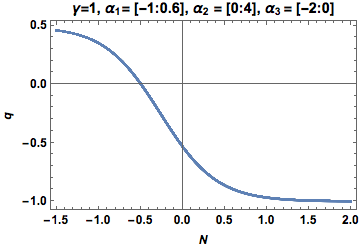} 
\caption{Plot of deceleration parameter against $N = \ln a$ from the fixed point a.}

\end{figure}

\begin{figure}[H]
\centering
\includegraphics[scale=0.6]{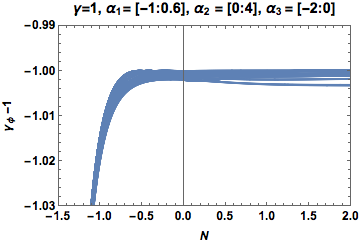} 
\caption{Plot of equation of state parameter $w_{\phi} = \gamma_{\phi} -1$ against $N = \ln a$ from the fixed point a.}
\end{figure}

\section{Discussions}

In this work, we have performed dynamical systems analysis of phantom dark energy models with five different potentials. The study of these potentials leads to a general parametrization of the potential function $\Gamma$. This general parametrization is not only valid for these five potentials but also applicable to a class of potentials for which $f$ can be written in our generalized form. We have tried to keep the analysis as much general as possible. Stability and the cosmological behaviours of the fixed point $a$ are independent of the choice of the potentials and it deserves to be the best choice as the beginning of the universe, whereas the stability of other fixed points depends on the choice of potentials but their cosmological behaviour is generic. Each one of them is dark energy dominated and decelerated.  So every potential which follows our general parametrization scheme for phantom dark energy models has the same beginning and same ultimate fate.

We have performed a detail numerical analysis of the system. As the fixed point $a$ is the most favorable to the beginning of the universe we have allowed the system to evolve from the surrounding of this fixed points. In this numerical analysis, the parameter range has been chosen in such a way that includes all the potentials in Table I. However this numerical analysis is not only restricted to these five potentials. The solutions around the fixed point $a$ evolved to the fixed point $b$. The numerical solutions show that the universe started from a matter dominated, decelerated saddle point and very recently around $z \simeq 0.55 $. It smoothly transits from the decelerated expansion state to an accelerated state. It also shows that the universe is presently dark energy dominated as $\Omega_{\phi} \simeq 0.67$ and the equation of state of the scalar field is $w_{\phi} < -1$. All these above findings agree well with the current cosmological observations and dynamical systems analysis of the system. 

This analysis does not show any favor to a particular form of the potential. It shows that a class potentials are allowed to describe the accelerated expansion of the universe. So the arbitrariness of the choice of potentials remains the same.

Though this analysis is done by choosing five different potentials one can consider more potentials to do the same. We have restricted ourself to these five examples so that the analysis does not become unnecessarily long. It is also interesting to note that this general parametrization of $f$ does not depend on the particular scalar field dark energy model; rather it comes from the definition of $\Gamma$ and $\lambda$. So this general parametrization of $f$ will also be valid for other scalar field dark energy models like quintessence, quintom etc where the dynamical system's variables can be considered in similar fashion.

\begin{acknowledgments}
N.R. acknowledges PRODEP for financial support. This work was partially supported by Programa para el Desarrollo Profesional Docente; Direcci\'on de Apoyo a la Investigaci\'on y al Posgrado, Universidad de Guanajuato, research Grants No. 732/2017 y 878/2017; CONACyT M\'exico under Grants No. 167335, No. 179881, No. 269652 and Fronteras 281; and the Fundaci\'on Marcos Moshinsky.
\end{acknowledgments}

\bibliography{phantom}

\begin{thebibliography}{48}%
\makeatletter
\providecommand \@ifxundefined [1]{%
 \@ifx{#1\undefined}
}%
\providecommand \@ifnum [1]{%
 \ifnum #1\expandafter \@firstoftwo
 \else \expandafter \@secondoftwo
 \fi
}%
\providecommand \@ifx [1]{%
 \ifx #1\expandafter \@firstoftwo
 \else \expandafter \@secondoftwo
 \fi
}%
\providecommand \natexlab [1]{#1}%
\providecommand \enquote  [1]{``#1''}%
\providecommand \bibnamefont  [1]{#1}%
\providecommand \bibfnamefont [1]{#1}%
\providecommand \citenamefont [1]{#1}%
\providecommand \href@noop [0]{\@secondoftwo}%
\providecommand \href [0]{\begingroup \@sanitize@url \@href}%
\providecommand \@href[1]{\@@startlink{#1}\@@href}%
\providecommand \@@href[1]{\endgroup#1\@@endlink}%
\providecommand \@sanitize@url [0]{\catcode `\\12\catcode `\$12\catcode
  `\&12\catcode `\#12\catcode `\^12\catcode `\_12\catcode `\%12\relax}%
\providecommand \@@startlink[1]{}%
\providecommand \@@endlink[0]{}%
\providecommand \url  [0]{\begingroup\@sanitize@url \@url }%
\providecommand \@url [1]{\endgroup\@href {#1}{\urlprefix }}%
\providecommand \urlprefix  [0]{URL }%
\providecommand \Eprint [0]{\href }%
\providecommand \doibase [0]{http://dx.doi.org/}%
\providecommand \selectlanguage [0]{\@gobble}%
\providecommand \bibinfo  [0]{\@secondoftwo}%
\providecommand \bibfield  [0]{\@secondoftwo}%
\providecommand \translation [1]{[#1]}%
\providecommand \BibitemOpen [0]{}%
\providecommand \bibitemStop [0]{}%
\providecommand \bibitemNoStop [0]{.\EOS\space}%
\providecommand \EOS [0]{\spacefactor3000\relax}%
\providecommand \BibitemShut  [1]{\csname bibitem#1\endcsname}%
\let\auto@bib@innerbib\@empty
%</preamble>
\bibitem [{\citenamefont {Ade}\ \emph {et~al.}(2016{\natexlab{a}})\citenamefont
  {Ade}, \citenamefont {Aghanim}, \citenamefont {Arnaud}, \citenamefont
  {Ashdown}, \citenamefont {Aumont}, \citenamefont {Baccigalupi}, \citenamefont
  {Banday}, \citenamefont {Barreiro}, \citenamefont {Bartlett}, \citenamefont
  {Bartolo} \emph {et~al.}}]{ade2016planck}%
  \BibitemOpen
  \bibfield  {author} {\bibinfo {author} {\bibfnamefont {P.}~\bibnamefont
  {Ade}}, \bibinfo {author} {\bibfnamefont {N.}~\bibnamefont {Aghanim}},
  \bibinfo {author} {\bibfnamefont {M.}~\bibnamefont {Arnaud}}, \bibinfo
  {author} {\bibfnamefont {M.}~\bibnamefont {Ashdown}}, \bibinfo {author}
  {\bibfnamefont {J.}~\bibnamefont {Aumont}}, \bibinfo {author} {\bibfnamefont
  {C.}~\bibnamefont {Baccigalupi}}, \bibinfo {author} {\bibfnamefont
  {A.}~\bibnamefont {Banday}}, \bibinfo {author} {\bibfnamefont
  {R.}~\bibnamefont {Barreiro}}, \bibinfo {author} {\bibfnamefont
  {J.}~\bibnamefont {Bartlett}}, \bibinfo {author} {\bibfnamefont
  {N.}~\bibnamefont {Bartolo}},  \emph {et~al.},\ }\href@noop {} {\bibfield
  {journal} {\bibinfo  {journal} {Astronomy \& Astrophysics}\ }\textbf
  {\bibinfo {volume} {594}},\ \bibinfo {pages} {A13} (\bibinfo {year}
  {2016}{\natexlab{a}})}\BibitemShut {NoStop}%
\bibitem [{\citenamefont {Perlmutter}\ \emph {et~al.}(1999)\citenamefont
  {Perlmutter}, \citenamefont {Collaboration} \emph
  {et~al.}}]{perlmutter1999astrophys}%
  \BibitemOpen
  \bibfield  {author} {\bibinfo {author} {\bibfnamefont {S.}~\bibnamefont
  {Perlmutter}}, \bibinfo {author} {\bibfnamefont {S.~C.~P.}\ \bibnamefont
  {Collaboration}},  \emph {et~al.},\ }\href@noop {} {\bibfield  {journal}
  {\bibinfo  {journal} {Astron. J}\ }\textbf {\bibinfo {volume} {116}},\
  \bibinfo {pages} {1009} (\bibinfo {year} {1999})}\BibitemShut {NoStop}%
\bibitem [{\citenamefont {Riess}\ \emph {et~al.}(1998)\citenamefont {Riess},
  \citenamefont {Filippenko}, \citenamefont {Challis}, \citenamefont
  {Clocchiatti}, \citenamefont {Diercks}, \citenamefont {Garnavich},
  \citenamefont {Gilliland}, \citenamefont {Hogan}, \citenamefont {Jha},
  \citenamefont {Kirshner} \emph {et~al.}}]{riess1998observational}%
  \BibitemOpen
  \bibfield  {author} {\bibinfo {author} {\bibfnamefont {A.~G.}\ \bibnamefont
  {Riess}}, \bibinfo {author} {\bibfnamefont {A.~V.}\ \bibnamefont
  {Filippenko}}, \bibinfo {author} {\bibfnamefont {P.}~\bibnamefont {Challis}},
  \bibinfo {author} {\bibfnamefont {A.}~\bibnamefont {Clocchiatti}}, \bibinfo
  {author} {\bibfnamefont {A.}~\bibnamefont {Diercks}}, \bibinfo {author}
  {\bibfnamefont {P.~M.}\ \bibnamefont {Garnavich}}, \bibinfo {author}
  {\bibfnamefont {R.~L.}\ \bibnamefont {Gilliland}}, \bibinfo {author}
  {\bibfnamefont {C.~J.}\ \bibnamefont {Hogan}}, \bibinfo {author}
  {\bibfnamefont {S.}~\bibnamefont {Jha}}, \bibinfo {author} {\bibfnamefont
  {R.~P.}\ \bibnamefont {Kirshner}},  \emph {et~al.},\ }\href@noop {}
  {\bibfield  {journal} {\bibinfo  {journal} {The Astronomical Journal}\
  }\textbf {\bibinfo {volume} {116}},\ \bibinfo {pages} {1009} (\bibinfo {year}
  {1998})}\BibitemShut {NoStop}%
\bibitem [{\citenamefont {Hinshaw}\ \emph {et~al.}(2007)\citenamefont
  {Hinshaw}, \citenamefont {Nolta}, \citenamefont {Bennett}, \citenamefont
  {Bean}, \citenamefont {Dor{\'e}}, \citenamefont {Greason}, \citenamefont
  {Halpern}, \citenamefont {Hill}, \citenamefont {Jarosik}, \citenamefont
  {Kogut} \emph {et~al.}}]{hinshaw2007three}%
  \BibitemOpen
  \bibfield  {author} {\bibinfo {author} {\bibfnamefont {G.}~\bibnamefont
  {Hinshaw}}, \bibinfo {author} {\bibfnamefont {M.}~\bibnamefont {Nolta}},
  \bibinfo {author} {\bibfnamefont {C.}~\bibnamefont {Bennett}}, \bibinfo
  {author} {\bibfnamefont {R.}~\bibnamefont {Bean}}, \bibinfo {author}
  {\bibfnamefont {O.}~\bibnamefont {Dor{\'e}}}, \bibinfo {author}
  {\bibfnamefont {M.}~\bibnamefont {Greason}}, \bibinfo {author} {\bibfnamefont
  {M.}~\bibnamefont {Halpern}}, \bibinfo {author} {\bibfnamefont
  {R.}~\bibnamefont {Hill}}, \bibinfo {author} {\bibfnamefont {N.}~\bibnamefont
  {Jarosik}}, \bibinfo {author} {\bibfnamefont {A.}~\bibnamefont {Kogut}},
  \emph {et~al.},\ }\href@noop {} {\bibfield  {journal} {\bibinfo  {journal}
  {The Astrophysical Journal Supplement Series}\ }\textbf {\bibinfo {volume}
  {170}},\ \bibinfo {pages} {288} (\bibinfo {year} {2007})}\BibitemShut
  {NoStop}%
\bibitem [{\citenamefont {Tegmark}\ \emph {et~al.}(2004)\citenamefont
  {Tegmark}, \citenamefont {Strauss}, \citenamefont {Blanton}, \citenamefont
  {Abazajian}, \citenamefont {Dodelson}, \citenamefont {Sandvik}, \citenamefont
  {Wang}, \citenamefont {Weinberg}, \citenamefont {Zehavi}, \citenamefont
  {Bahcall} \emph {et~al.}}]{tegmark2004cosmological}%
  \BibitemOpen
  \bibfield  {author} {\bibinfo {author} {\bibfnamefont {M.}~\bibnamefont
  {Tegmark}}, \bibinfo {author} {\bibfnamefont {M.~A.}\ \bibnamefont
  {Strauss}}, \bibinfo {author} {\bibfnamefont {M.~R.}\ \bibnamefont
  {Blanton}}, \bibinfo {author} {\bibfnamefont {K.}~\bibnamefont {Abazajian}},
  \bibinfo {author} {\bibfnamefont {S.}~\bibnamefont {Dodelson}}, \bibinfo
  {author} {\bibfnamefont {H.}~\bibnamefont {Sandvik}}, \bibinfo {author}
  {\bibfnamefont {X.}~\bibnamefont {Wang}}, \bibinfo {author} {\bibfnamefont
  {D.~H.}\ \bibnamefont {Weinberg}}, \bibinfo {author} {\bibfnamefont
  {I.}~\bibnamefont {Zehavi}}, \bibinfo {author} {\bibfnamefont {N.~A.}\
  \bibnamefont {Bahcall}},  \emph {et~al.},\ }\href@noop {} {\bibfield
  {journal} {\bibinfo  {journal} {Physical Review D}\ }\textbf {\bibinfo
  {volume} {69}},\ \bibinfo {pages} {103501} (\bibinfo {year}
  {2004})}\BibitemShut {NoStop}%
\bibitem [{\citenamefont {Seljak}\ \emph {et~al.}(2005)\citenamefont {Seljak},
  \citenamefont {Makarov}, \citenamefont {McDonald}, \citenamefont {Anderson},
  \citenamefont {Bahcall}, \citenamefont {Brinkmann}, \citenamefont {Burles},
  \citenamefont {Cen}, \citenamefont {Doi}, \citenamefont {Gunn} \emph
  {et~al.}}]{seljak2005cosmological}%
  \BibitemOpen
  \bibfield  {author} {\bibinfo {author} {\bibfnamefont {U.}~\bibnamefont
  {Seljak}}, \bibinfo {author} {\bibfnamefont {A.}~\bibnamefont {Makarov}},
  \bibinfo {author} {\bibfnamefont {P.}~\bibnamefont {McDonald}}, \bibinfo
  {author} {\bibfnamefont {S.~F.}\ \bibnamefont {Anderson}}, \bibinfo {author}
  {\bibfnamefont {N.~A.}\ \bibnamefont {Bahcall}}, \bibinfo {author}
  {\bibfnamefont {J.}~\bibnamefont {Brinkmann}}, \bibinfo {author}
  {\bibfnamefont {S.}~\bibnamefont {Burles}}, \bibinfo {author} {\bibfnamefont
  {R.}~\bibnamefont {Cen}}, \bibinfo {author} {\bibfnamefont {M.}~\bibnamefont
  {Doi}}, \bibinfo {author} {\bibfnamefont {J.~E.}\ \bibnamefont {Gunn}},
  \emph {et~al.},\ }\href@noop {} {\bibfield  {journal} {\bibinfo  {journal}
  {Physical Review D}\ }\textbf {\bibinfo {volume} {71}},\ \bibinfo {pages}
  {103515} (\bibinfo {year} {2005})}\BibitemShut {NoStop}%
\bibitem [{\citenamefont {Padmanabhan}(2003)}]{padmanabhan2003cosmological}%
  \BibitemOpen
  \bibfield  {author} {\bibinfo {author} {\bibfnamefont {T.}~\bibnamefont
  {Padmanabhan}},\ }\href@noop {} {\bibfield  {journal} {\bibinfo  {journal}
  {Physics Reports}\ }\textbf {\bibinfo {volume} {380}},\ \bibinfo {pages}
  {235} (\bibinfo {year} {2003})}\BibitemShut {NoStop}%
\bibitem [{\citenamefont {Weinberg}(1989)}]{weinberg1989cosmological}%
  \BibitemOpen
  \bibfield  {author} {\bibinfo {author} {\bibfnamefont {S.}~\bibnamefont
  {Weinberg}},\ }\href@noop {} {\bibfield  {journal} {\bibinfo  {journal}
  {Reviews of modern physics}\ }\textbf {\bibinfo {volume} {61}},\ \bibinfo
  {pages} {1} (\bibinfo {year} {1989})}\BibitemShut {NoStop}%
\bibitem [{\citenamefont {Copeland}\ \emph {et~al.}(2006)\citenamefont
  {Copeland}, \citenamefont {Sami},\ and\ \citenamefont
  {Tsujikawa}}]{copeland2006dynamics}%
  \BibitemOpen
  \bibfield  {author} {\bibinfo {author} {\bibfnamefont {E.~J.}\ \bibnamefont
  {Copeland}}, \bibinfo {author} {\bibfnamefont {M.}~\bibnamefont {Sami}}, \
  and\ \bibinfo {author} {\bibfnamefont {S.}~\bibnamefont {Tsujikawa}},\
  }\href@noop {} {\bibfield  {journal} {\bibinfo  {journal} {International
  Journal of Modern Physics D}\ }\textbf {\bibinfo {volume} {15}},\ \bibinfo
  {pages} {1753} (\bibinfo {year} {2006})}\BibitemShut {NoStop}%
\bibitem [{\citenamefont {Caldwell}\ \emph {et~al.}(1998)\citenamefont
  {Caldwell}, \citenamefont {Dave},\ and\ \citenamefont
  {Steinhardt}}]{cladwell}%
  \BibitemOpen
  \bibfield  {author} {\bibinfo {author} {\bibfnamefont {R.~R.}\ \bibnamefont
  {Caldwell}}, \bibinfo {author} {\bibfnamefont {R.}~\bibnamefont {Dave}}, \
  and\ \bibinfo {author} {\bibfnamefont {P.~J.}\ \bibnamefont {Steinhardt}},\
  }\href {\doibase 10.1103/PhysRevLett.80.1582} {\bibfield  {journal} {\bibinfo
   {journal} {Phys. Rev. Lett.}\ }\textbf {\bibinfo {volume} {80}},\ \bibinfo
  {pages} {1582} (\bibinfo {year} {1998})}\BibitemShut {NoStop}%
\bibitem [{\citenamefont {Tsujikawa}(2013)}]{tsujikawa}%
  \BibitemOpen
  \bibfield  {author} {\bibinfo {author} {\bibfnamefont {S.}~\bibnamefont
  {Tsujikawa}},\ }\href@noop {} {\bibfield  {journal} {\bibinfo  {journal}
  {Classical and Quantum Gravity}\ }\textbf {\bibinfo {volume} {30}},\ \bibinfo
  {pages} {214003} (\bibinfo {year} {2013})}\BibitemShut {NoStop}%
\bibitem [{\citenamefont {Copeland}\ \emph {et~al.}(1998)\citenamefont
  {Copeland}, \citenamefont {Liddle},\ and\ \citenamefont
  {Wands}}]{copeland1998exponential}%
  \BibitemOpen
  \bibfield  {author} {\bibinfo {author} {\bibfnamefont {E.~J.}\ \bibnamefont
  {Copeland}}, \bibinfo {author} {\bibfnamefont {A.~R.}\ \bibnamefont
  {Liddle}}, \ and\ \bibinfo {author} {\bibfnamefont {D.}~\bibnamefont
  {Wands}},\ }\href@noop {} {\bibfield  {journal} {\bibinfo  {journal}
  {Physical Review D}\ }\textbf {\bibinfo {volume} {57}},\ \bibinfo {pages}
  {4686} (\bibinfo {year} {1998})}\BibitemShut {NoStop}%
\bibitem [{\citenamefont {Caldwell}(2002)}]{caldwell2002phantom}%
  \BibitemOpen
  \bibfield  {author} {\bibinfo {author} {\bibfnamefont {R.~R.}\ \bibnamefont
  {Caldwell}},\ }\href@noop {} {\bibfield  {journal} {\bibinfo  {journal}
  {Physics Letters B}\ }\textbf {\bibinfo {volume} {545}},\ \bibinfo {pages}
  {23} (\bibinfo {year} {2002})}\BibitemShut {NoStop}%
\bibitem [{\citenamefont {Caldwell}\ \emph {et~al.}()\citenamefont {Caldwell},
  \citenamefont {Kamionkowski},\ and\ \citenamefont
  {Weinberg}}]{caldwell91phantom}%
  \BibitemOpen
  \bibfield  {author} {\bibinfo {author} {\bibfnamefont {R.}~\bibnamefont
  {Caldwell}}, \bibinfo {author} {\bibfnamefont {M.}~\bibnamefont
  {Kamionkowski}}, \ and\ \bibinfo {author} {\bibfnamefont {N.}~\bibnamefont
  {Weinberg}},\ }\href@noop {} {\bibfield  {journal} {\bibinfo  {journal}
  {Phys. Rev. Lett}\ }\textbf {\bibinfo {volume} {91}},\ \bibinfo {pages}
  {071301}}\BibitemShut {NoStop}%
\bibitem [{\citenamefont {Ludwick}(2017)}]{ludwick2017viability}%
  \BibitemOpen
  \bibfield  {author} {\bibinfo {author} {\bibfnamefont {K.~J.}\ \bibnamefont
  {Ludwick}},\ }\href@noop {} {\bibfield  {journal} {\bibinfo  {journal}
  {Modern Physics Letters A}\ }\textbf {\bibinfo {volume} {32}},\ \bibinfo
  {pages} {1730025} (\bibinfo {year} {2017})}\BibitemShut {NoStop}%
\bibitem [{\citenamefont {Carroll}\ \emph {et~al.}(2003)\citenamefont
  {Carroll}, \citenamefont {Hoffman},\ and\ \citenamefont
  {Trodden}}]{carroll2003can}%
  \BibitemOpen
  \bibfield  {author} {\bibinfo {author} {\bibfnamefont {S.~M.}\ \bibnamefont
  {Carroll}}, \bibinfo {author} {\bibfnamefont {M.}~\bibnamefont {Hoffman}}, \
  and\ \bibinfo {author} {\bibfnamefont {M.}~\bibnamefont {Trodden}},\
  }\href@noop {} {\bibfield  {journal} {\bibinfo  {journal} {Physical Review
  D}\ }\textbf {\bibinfo {volume} {68}},\ \bibinfo {pages} {023509} (\bibinfo
  {year} {2003})}\BibitemShut {NoStop}%
\bibitem [{\citenamefont {Cline}\ \emph {et~al.}(2004)\citenamefont {Cline},
  \citenamefont {Jeon},\ and\ \citenamefont {Moore}}]{cline2004phantom}%
  \BibitemOpen
  \bibfield  {author} {\bibinfo {author} {\bibfnamefont {J.~M.}\ \bibnamefont
  {Cline}}, \bibinfo {author} {\bibfnamefont {S.}~\bibnamefont {Jeon}}, \ and\
  \bibinfo {author} {\bibfnamefont {G.~D.}\ \bibnamefont {Moore}},\ }\href@noop
  {} {\bibfield  {journal} {\bibinfo  {journal} {Physical Review D}\ }\textbf
  {\bibinfo {volume} {70}},\ \bibinfo {pages} {043543} (\bibinfo {year}
  {2004})}\BibitemShut {NoStop}%
\bibitem [{\citenamefont {Singh}\ \emph {et~al.}(2003)\citenamefont {Singh},
  \citenamefont {Sami},\ and\ \citenamefont {Dadhich}}]{singh2003cosmological}%
  \BibitemOpen
  \bibfield  {author} {\bibinfo {author} {\bibfnamefont {P.}~\bibnamefont
  {Singh}}, \bibinfo {author} {\bibfnamefont {M.}~\bibnamefont {Sami}}, \ and\
  \bibinfo {author} {\bibfnamefont {N.}~\bibnamefont {Dadhich}},\ }\href@noop
  {} {\bibfield  {journal} {\bibinfo  {journal} {Physical Review D}\ }\textbf
  {\bibinfo {volume} {68}},\ \bibinfo {pages} {023522} (\bibinfo {year}
  {2003})}\BibitemShut {NoStop}%
\bibitem [{\citenamefont {Sami}\ and\ \citenamefont
  {Toporensky}(2004)}]{sami2004phantom}%
  \BibitemOpen
  \bibfield  {author} {\bibinfo {author} {\bibfnamefont {M.}~\bibnamefont
  {Sami}}\ and\ \bibinfo {author} {\bibfnamefont {A.}~\bibnamefont
  {Toporensky}},\ }\href@noop {} {\bibfield  {journal} {\bibinfo  {journal}
  {Modern Physics Letters A}\ }\textbf {\bibinfo {volume} {19}},\ \bibinfo
  {pages} {1509} (\bibinfo {year} {2004})}\BibitemShut {NoStop}%
\bibitem [{\citenamefont {Zlatev}\ \emph {et~al.}(1999)\citenamefont {Zlatev},
  \citenamefont {Wang},\ and\ \citenamefont
  {Steinhardt}}]{zlatev1999quintessence}%
  \BibitemOpen
  \bibfield  {author} {\bibinfo {author} {\bibfnamefont {I.}~\bibnamefont
  {Zlatev}}, \bibinfo {author} {\bibfnamefont {L.}~\bibnamefont {Wang}}, \ and\
  \bibinfo {author} {\bibfnamefont {P.~J.}\ \bibnamefont {Steinhardt}},\
  }\href@noop {} {\bibfield  {journal} {\bibinfo  {journal} {Physical Review
  Letters}\ }\textbf {\bibinfo {volume} {82}},\ \bibinfo {pages} {896}
  (\bibinfo {year} {1999})}\BibitemShut {NoStop}%
\bibitem [{\citenamefont {De~La~Macorra}\ and\ \citenamefont
  {Piccinelli}(2000)}]{de2000cosmological}%
  \BibitemOpen
  \bibfield  {author} {\bibinfo {author} {\bibfnamefont {A.}~\bibnamefont
  {De~La~Macorra}}\ and\ \bibinfo {author} {\bibfnamefont {G.}~\bibnamefont
  {Piccinelli}},\ }\href@noop {} {\bibfield  {journal} {\bibinfo  {journal}
  {Physical Review D}\ }\textbf {\bibinfo {volume} {61}},\ \bibinfo {pages}
  {123503} (\bibinfo {year} {2000})}\BibitemShut {NoStop}%
\bibitem [{\citenamefont {Ng}\ \emph {et~al.}(2001)\citenamefont {Ng},
  \citenamefont {Nunes},\ and\ \citenamefont {Rosati}}]{ng2001applications}%
  \BibitemOpen
  \bibfield  {author} {\bibinfo {author} {\bibfnamefont {S.}~\bibnamefont
  {Ng}}, \bibinfo {author} {\bibfnamefont {N.}~\bibnamefont {Nunes}}, \ and\
  \bibinfo {author} {\bibfnamefont {F.}~\bibnamefont {Rosati}},\ }\href@noop {}
  {\bibfield  {journal} {\bibinfo  {journal} {Physical Review D}\ }\textbf
  {\bibinfo {volume} {64}},\ \bibinfo {pages} {083510} (\bibinfo {year}
  {2001})}\BibitemShut {NoStop}%
\bibitem [{\citenamefont {Corasaniti}\ and\ \citenamefont
  {Copeland}(2003)}]{corasaniti2003model}%
  \BibitemOpen
  \bibfield  {author} {\bibinfo {author} {\bibfnamefont {P.~S.}\ \bibnamefont
  {Corasaniti}}\ and\ \bibinfo {author} {\bibfnamefont {E.}~\bibnamefont
  {Copeland}},\ }\href@noop {} {\bibfield  {journal} {\bibinfo  {journal}
  {Physical Review D}\ }\textbf {\bibinfo {volume} {67}},\ \bibinfo {pages}
  {063521} (\bibinfo {year} {2003})}\BibitemShut {NoStop}%
\bibitem [{\citenamefont {Caldwell}\ and\ \citenamefont
  {Linder}(2005)}]{caldwell2005limits}%
  \BibitemOpen
  \bibfield  {author} {\bibinfo {author} {\bibfnamefont {R.}~\bibnamefont
  {Caldwell}}\ and\ \bibinfo {author} {\bibfnamefont {E.~V.}\ \bibnamefont
  {Linder}},\ }\href@noop {} {\bibfield  {journal} {\bibinfo  {journal}
  {Physical review letters}\ }\textbf {\bibinfo {volume} {95}},\ \bibinfo
  {pages} {141301} (\bibinfo {year} {2005})}\BibitemShut {NoStop}%
\bibitem [{\citenamefont {Linder}(2006)}]{linder2006paths}%
  \BibitemOpen
  \bibfield  {author} {\bibinfo {author} {\bibfnamefont {E.~V.}\ \bibnamefont
  {Linder}},\ }\href@noop {} {\bibfield  {journal} {\bibinfo  {journal}
  {Physical Review D}\ }\textbf {\bibinfo {volume} {73}},\ \bibinfo {pages}
  {063010} (\bibinfo {year} {2006})}\BibitemShut {NoStop}%
\bibitem [{\citenamefont {Scherrer}\ and\ \citenamefont
  {Sen}(2008)}]{scherrer2008thawing}%
  \BibitemOpen
  \bibfield  {author} {\bibinfo {author} {\bibfnamefont {R.~J.}\ \bibnamefont
  {Scherrer}}\ and\ \bibinfo {author} {\bibfnamefont {A.}~\bibnamefont {Sen}},\
  }\href@noop {} {\bibfield  {journal} {\bibinfo  {journal} {Physical Review
  D}\ }\textbf {\bibinfo {volume} {77}},\ \bibinfo {pages} {083515} (\bibinfo
  {year} {2008})}\BibitemShut {NoStop}%
\bibitem [{\citenamefont {Chiba}(2009)}]{chiba2009slow}%
  \BibitemOpen
  \bibfield  {author} {\bibinfo {author} {\bibfnamefont {T.}~\bibnamefont
  {Chiba}},\ }\href@noop {} {\bibfield  {journal} {\bibinfo  {journal}
  {Physical Review D}\ }\textbf {\bibinfo {volume} {79}},\ \bibinfo {pages}
  {083517} (\bibinfo {year} {2009})}\BibitemShut {NoStop}%
\bibitem [{\citenamefont {Sahni}(2004)}]{sahni20045}%
  \BibitemOpen
  \bibfield  {author} {\bibinfo {author} {\bibfnamefont {V.}~\bibnamefont
  {Sahni}},\ }in\ \href@noop {} {\emph {\bibinfo {booktitle} {The Physics of
  the Early Universe}}}\ (\bibinfo  {publisher} {Springer},\ \bibinfo {year}
  {2004})\ pp.\ \bibinfo {pages} {141--179}\BibitemShut {NoStop}%
\bibitem [{\citenamefont {Roy}\ and\ \citenamefont
  {Banerjee}(2014{\natexlab{a}})}]{roy2014quintessence}%
  \BibitemOpen
  \bibfield  {author} {\bibinfo {author} {\bibfnamefont {N.}~\bibnamefont
  {Roy}}\ and\ \bibinfo {author} {\bibfnamefont {N.}~\bibnamefont {Banerjee}},\
  }\href@noop {} {\bibfield  {journal} {\bibinfo  {journal} {The European
  Physical Journal Plus}\ }\textbf {\bibinfo {volume} {129}},\ \bibinfo {pages}
  {162} (\bibinfo {year} {2014}{\natexlab{a}})}\BibitemShut {NoStop}%
\bibitem [{\citenamefont {Roy}\ and\ \citenamefont
  {Banerjee}(2014{\natexlab{b}})}]{roy2014tracking}%
  \BibitemOpen
  \bibfield  {author} {\bibinfo {author} {\bibfnamefont {N.}~\bibnamefont
  {Roy}}\ and\ \bibinfo {author} {\bibfnamefont {N.}~\bibnamefont {Banerjee}},\
  }\href@noop {} {\bibfield  {journal} {\bibinfo  {journal} {General Relativity
  and Gravitation}\ }\textbf {\bibinfo {volume} {46}},\ \bibinfo {pages} {1651}
  (\bibinfo {year} {2014}{\natexlab{b}})}\BibitemShut {NoStop}%
\bibitem [{\citenamefont {Roy}\ and\ \citenamefont
  {Banerjee}(2015)}]{roy2015dynamical}%
  \BibitemOpen
  \bibfield  {author} {\bibinfo {author} {\bibfnamefont {N.}~\bibnamefont
  {Roy}}\ and\ \bibinfo {author} {\bibfnamefont {N.}~\bibnamefont {Banerjee}},\
  }\href@noop {} {\bibfield  {journal} {\bibinfo  {journal} {Annals of
  Physics}\ }\textbf {\bibinfo {volume} {356}},\ \bibinfo {pages} {452}
  (\bibinfo {year} {2015})}\BibitemShut {NoStop}%
\bibitem [{\citenamefont {Dutta}\ \emph {et~al.}(2016)\citenamefont {Dutta},
  \citenamefont {Khyllep},\ and\ \citenamefont
  {Tamanini}}]{dutta2016cosmological}%
  \BibitemOpen
  \bibfield  {author} {\bibinfo {author} {\bibfnamefont {J.}~\bibnamefont
  {Dutta}}, \bibinfo {author} {\bibfnamefont {W.}~\bibnamefont {Khyllep}}, \
  and\ \bibinfo {author} {\bibfnamefont {N.}~\bibnamefont {Tamanini}},\
  }\href@noop {} {\bibfield  {journal} {\bibinfo  {journal} {Physical Review
  D}\ }\textbf {\bibinfo {volume} {93}},\ \bibinfo {pages} {063004} (\bibinfo
  {year} {2016})}\BibitemShut {NoStop}%
\bibitem [{\citenamefont {Bhatia}\ and\ \citenamefont
  {Sur}(2017)}]{Bhatia:2017wdh}%
  \BibitemOpen
  \bibfield  {author} {\bibinfo {author} {\bibfnamefont {A.~S.}\ \bibnamefont
  {Bhatia}}\ and\ \bibinfo {author} {\bibfnamefont {S.}~\bibnamefont {Sur}},\
  }\href {\doibase 10.1142/S0218271817501498} {\bibfield  {journal} {\bibinfo
  {journal} {Int. J. Mod. Phys.}\ }\textbf {\bibinfo {volume} {D26}},\ \bibinfo
  {pages} {1750149} (\bibinfo {year} {2017})},\ \Eprint
  {http://arxiv.org/abs/1702.01267} {arXiv:1702.01267 [gr-qc]} \BibitemShut
  {NoStop}%
%%CITATION = ARXIV:1702.01267;%%
\bibitem [{\citenamefont {Sola}\ \emph {et~al.}(2017)\citenamefont {Sola},
  \citenamefont {Gomez-Valent},\ and\ \citenamefont
  {de~Cruz~Pérez}}]{Sola:2016hnq}%
  \BibitemOpen
  \bibfield  {author} {\bibinfo {author} {\bibfnamefont {J.}~\bibnamefont
  {Sola}}, \bibinfo {author} {\bibfnamefont {A.}~\bibnamefont {Gomez-Valent}},
  \ and\ \bibinfo {author} {\bibfnamefont {J.}~\bibnamefont {de~Cruz~Pérez}},\
  }\href {\doibase 10.1142/S0217732317500547} {\bibfield  {journal} {\bibinfo
  {journal} {Mod. Phys. Lett.}\ }\textbf {\bibinfo {volume} {A32}},\ \bibinfo
  {pages} {1750054} (\bibinfo {year} {2017})},\ \Eprint
  {http://arxiv.org/abs/1610.08965} {arXiv:1610.08965 [astro-ph.CO]}
  \BibitemShut {NoStop}%
%%CITATION = ARXIV:1610.08965;%%
\bibitem [{\citenamefont {Wainwright}\ and\ \citenamefont
  {Ellis}(2005)}]{wainwright2005dynamical}%
  \BibitemOpen
  \bibfield  {author} {\bibinfo {author} {\bibfnamefont {J.}~\bibnamefont
  {Wainwright}}\ and\ \bibinfo {author} {\bibfnamefont {G.~F.~R.}\ \bibnamefont
  {Ellis}},\ }\href@noop {} {\emph {\bibinfo {title} {Dynamical systems in
  cosmology}}}\ (\bibinfo  {publisher} {Cambridge University Press},\ \bibinfo
  {year} {2005})\BibitemShut {NoStop}%
\bibitem [{\citenamefont {Coley}(2013)}]{coley2013dynamical}%
  \BibitemOpen
  \bibfield  {author} {\bibinfo {author} {\bibfnamefont {A.~A.}\ \bibnamefont
  {Coley}},\ }\href@noop {} {\emph {\bibinfo {title} {Dynamical systems and
  cosmology}}},\ Vol.\ \bibinfo {volume} {291}\ (\bibinfo  {publisher}
  {Springer Science \& Business Media},\ \bibinfo {year} {2013})\BibitemShut
  {NoStop}%
\bibitem [{\citenamefont {Bahamonde}\ \emph {et~al.}(2017)\citenamefont
  {Bahamonde}, \citenamefont {Boehmer}, \citenamefont {Carloni}, \citenamefont
  {Copeland}, \citenamefont {Fang},\ and\ \citenamefont
  {Tamanini}}]{bahamonde2017dynamical}%
  \BibitemOpen
  \bibfield  {author} {\bibinfo {author} {\bibfnamefont {S.}~\bibnamefont
  {Bahamonde}}, \bibinfo {author} {\bibfnamefont {C.~G.}\ \bibnamefont
  {Boehmer}}, \bibinfo {author} {\bibfnamefont {S.}~\bibnamefont {Carloni}},
  \bibinfo {author} {\bibfnamefont {E.~J.}\ \bibnamefont {Copeland}}, \bibinfo
  {author} {\bibfnamefont {W.}~\bibnamefont {Fang}}, \ and\ \bibinfo {author}
  {\bibfnamefont {N.}~\bibnamefont {Tamanini}},\ }\href@noop {} {\bibfield
  {journal} {\bibinfo  {journal} {arXiv preprint arXiv:1712.03107}\ } (\bibinfo
  {year} {2017})}\BibitemShut {NoStop}%
\bibitem [{\citenamefont {Guo}\ \emph {et~al.}(2004)\citenamefont {Guo},
  \citenamefont {Piao},\ and\ \citenamefont {Zhang}}]{guo2004attractor}%
  \BibitemOpen
  \bibfield  {author} {\bibinfo {author} {\bibfnamefont {Z.-K.}\ \bibnamefont
  {Guo}}, \bibinfo {author} {\bibfnamefont {Y.-S.}\ \bibnamefont {Piao}}, \
  and\ \bibinfo {author} {\bibfnamefont {Y.-Z.}\ \bibnamefont {Zhang}},\
  }\href@noop {} {\bibfield  {journal} {\bibinfo  {journal} {Physics Letters
  B}\ }\textbf {\bibinfo {volume} {594}},\ \bibinfo {pages} {247} (\bibinfo
  {year} {2004})}\BibitemShut {NoStop}%
\bibitem [{\citenamefont {Urena-Lopez}(2005)}]{urena2005scalar}%
  \BibitemOpen
  \bibfield  {author} {\bibinfo {author} {\bibfnamefont {L.~A.}\ \bibnamefont
  {Urena-Lopez}},\ }\href@noop {} {\bibfield  {journal} {\bibinfo  {journal}
  {Journal of Cosmology and Astroparticle Physics}\ }\textbf {\bibinfo {volume}
  {2005}},\ \bibinfo {pages} {013} (\bibinfo {year} {2005})}\BibitemShut
  {NoStop}%
\bibitem [{\citenamefont {Mahata}\ and\ \citenamefont
  {Chakraborty}(2014)}]{mahata2014dynamical}%
  \BibitemOpen
  \bibfield  {author} {\bibinfo {author} {\bibfnamefont {N.}~\bibnamefont
  {Mahata}}\ and\ \bibinfo {author} {\bibfnamefont {S.}~\bibnamefont
  {Chakraborty}},\ }\href@noop {} {\bibfield  {journal} {\bibinfo  {journal}
  {General Relativity and Gravitation}\ }\textbf {\bibinfo {volume} {46}},\
  \bibinfo {pages} {1721} (\bibinfo {year} {2014})}\BibitemShut {NoStop}%
\bibitem [{\citenamefont {Fang}\ \emph {et~al.}(2009)\citenamefont {Fang},
  \citenamefont {Li}, \citenamefont {Zhang},\ and\ \citenamefont
  {Lu}}]{fang2009exact}%
  \BibitemOpen
  \bibfield  {author} {\bibinfo {author} {\bibfnamefont {W.}~\bibnamefont
  {Fang}}, \bibinfo {author} {\bibfnamefont {Y.}~\bibnamefont {Li}}, \bibinfo
  {author} {\bibfnamefont {K.}~\bibnamefont {Zhang}}, \ and\ \bibinfo {author}
  {\bibfnamefont {H.-Q.}\ \bibnamefont {Lu}},\ }\href@noop {} {\bibfield
  {journal} {\bibinfo  {journal} {Classical and Quantum Gravity}\ }\textbf
  {\bibinfo {volume} {26}},\ \bibinfo {pages} {155005} (\bibinfo {year}
  {2009})}\BibitemShut {NoStop}%
\bibitem [{\citenamefont {Frieman}\ \emph {et~al.}(1995)\citenamefont
  {Frieman}, \citenamefont {Hill}, \citenamefont {Stebbins},\ and\
  \citenamefont {Waga}}]{frieman1995cosmology}%
  \BibitemOpen
  \bibfield  {author} {\bibinfo {author} {\bibfnamefont {J.~A.}\ \bibnamefont
  {Frieman}}, \bibinfo {author} {\bibfnamefont {C.~T.}\ \bibnamefont {Hill}},
  \bibinfo {author} {\bibfnamefont {A.}~\bibnamefont {Stebbins}}, \ and\
  \bibinfo {author} {\bibfnamefont {I.}~\bibnamefont {Waga}},\ }\href@noop {}
  {\bibfield  {journal} {\bibinfo  {journal} {Physical Review Letters}\
  }\textbf {\bibinfo {volume} {75}},\ \bibinfo {pages} {2077} (\bibinfo {year}
  {1995})}\BibitemShut {NoStop}%
\bibitem [{\citenamefont {Linde}(1983)}]{linde1983chaotic}%
  \BibitemOpen
  \bibfield  {author} {\bibinfo {author} {\bibfnamefont {A.~D.}\ \bibnamefont
  {Linde}},\ }\href@noop {} {\bibfield  {journal} {\bibinfo  {journal} {Physics
  Letters B}\ }\textbf {\bibinfo {volume} {129}},\ \bibinfo {pages} {177}
  (\bibinfo {year} {1983})}\BibitemShut {NoStop}%
\bibitem [{\citenamefont {Sahni}\ and\ \citenamefont
  {Wang}(2000)}]{sahni2000new}%
  \BibitemOpen
  \bibfield  {author} {\bibinfo {author} {\bibfnamefont {V.}~\bibnamefont
  {Sahni}}\ and\ \bibinfo {author} {\bibfnamefont {L.}~\bibnamefont {Wang}},\
  }\href@noop {} {\bibfield  {journal} {\bibinfo  {journal} {Physical Review
  D}\ }\textbf {\bibinfo {volume} {62}},\ \bibinfo {pages} {103517} (\bibinfo
  {year} {2000})}\BibitemShut {NoStop}%
\bibitem [{\citenamefont {Sahni}\ and\ \citenamefont
  {Starobinsky}(2000)}]{sahni2000case}%
  \BibitemOpen
  \bibfield  {author} {\bibinfo {author} {\bibfnamefont {V.}~\bibnamefont
  {Sahni}}\ and\ \bibinfo {author} {\bibfnamefont {A.}~\bibnamefont
  {Starobinsky}},\ }\href@noop {} {\bibfield  {journal} {\bibinfo  {journal}
  {International Journal of Modern Physics D}\ }\textbf {\bibinfo {volume}
  {9}},\ \bibinfo {pages} {373} (\bibinfo {year} {2000})}\BibitemShut {NoStop}%
\bibitem [{\citenamefont {Urena-Lopez}\ and\ \citenamefont
  {Matos}(2000)}]{urena2000new}%
  \BibitemOpen
  \bibfield  {author} {\bibinfo {author} {\bibfnamefont {L.~A.}\ \bibnamefont
  {Urena-Lopez}}\ and\ \bibinfo {author} {\bibfnamefont {T.}~\bibnamefont
  {Matos}},\ }\href@noop {} {\bibfield  {journal} {\bibinfo  {journal}
  {Physical Review D}\ }\textbf {\bibinfo {volume} {62}},\ \bibinfo {pages}
  {081302} (\bibinfo {year} {2000})}\BibitemShut {NoStop}%
\bibitem [{\citenamefont {Ure{\~n}a-L{\'o}pez}(2016)}]{urena2016new}%
  \BibitemOpen
  \bibfield  {author} {\bibinfo {author} {\bibfnamefont {L.~A.}\ \bibnamefont
  {Ure{\~n}a-L{\'o}pez}},\ }\href@noop {} {\bibfield  {journal} {\bibinfo
  {journal} {Physical Review D}\ }\textbf {\bibinfo {volume} {94}},\ \bibinfo
  {pages} {063532} (\bibinfo {year} {2016})}\BibitemShut {NoStop}%
\bibitem [{\citenamefont {Ade}\ \emph {et~al.}(2016{\natexlab{b}})\citenamefont
  {Ade} \emph {et~al.}}]{Ade:2015xua}%
  \BibitemOpen
  \bibfield  {author} {\bibinfo {author} {\bibfnamefont {P.~A.~R.}\
  \bibnamefont {Ade}} \emph {et~al.} (\bibinfo {collaboration} {Planck}),\
  }\href {\doibase 10.1051/0004-6361/201525830} {\bibfield  {journal} {\bibinfo
   {journal} {Astron. Astrophys.}\ }\textbf {\bibinfo {volume} {594}},\
  \bibinfo {pages} {A13} (\bibinfo {year} {2016}{\natexlab{b}})},\ \Eprint
  {http://arxiv.org/abs/1502.01589} {arXiv:1502.01589 [astro-ph.CO]}
  \BibitemShut {NoStop}%
%%CITATION = ARXIV:1502.01589;%%
\end{thebibliography}%

\newpage
\appendix*

 \begin{widetext}	
 	\section{Expression of the eigenvalues denoted by $m$ and $n$}

 	 \begin{eqnarray*}\nonumber
 	m_d = m_f &=& \frac{1}{4}  [4 \alpha _1 \lambda_- ^2 +2 \alpha _2 \lambda_- -\lambda_+  ^2-6\\
 	&&  +  \sqrt{(-4 \alpha _1 \lambda_- ^2-2 \alpha _2 \lambda_- +\lambda_+ ^2+6)^2-8 (-3 \alpha _1 \lambda_- ^4-2 \alpha _2 \lambda_- ^3-18 \alpha _1 \lambda_- ^2-\alpha _3 \lambda_- ^2-12 \alpha _2 \lambda_- -6 \alpha _3)}]
 	\end{eqnarray*}

 \begin{eqnarray*}\nonumber
n_d = n_f&=& \frac{1}{4}  [4 \alpha _1 \lambda_- ^2 +2 \alpha _2 \lambda_- -\lambda_+  ^2-6\\
&&  +  \sqrt{(-4 \alpha _1 \lambda_- ^2-2 \alpha _2 \lambda_- +\lambda_+ ^2+6)^2-8 (-3 \alpha _1 \lambda_- ^4-2 \alpha _2 \lambda_- ^3-18 \alpha _1 \lambda_- ^2-\alpha _3 \lambda_- ^2-12 \alpha _2 \lambda_- -6 \alpha _3)}]
\end{eqnarray*}

 \begin{eqnarray*}\nonumber
  m_e = m_g  &=&\frac{1}{4}  [4 \alpha _1 \lambda_+ ^2 +2 \alpha _2 \lambda_+ -\lambda_+  ^2-6\\
 &&  - \sqrt{(-4 \alpha _1 \lambda_+ ^2-2 \alpha _2 \lambda_+ +\lambda_+ ^2+6)^2-8 (-3 \alpha _1 \lambda_+ ^4-2 \alpha _2 \lambda_+ ^3-18 \alpha _1 \lambda_+ ^2-\alpha _3 \lambda_+ ^2-12 \alpha _2 \lambda_+ -6 \alpha _3)}  ~~ ]
 \end{eqnarray*}
 
 \begin{eqnarray*}\nonumber
  n_f = n_g&=& \frac{1}{4}  [4 \alpha _1 \lambda_+ ^2 +2 \alpha _2 \lambda_+ -\lambda_+  ^2-6\\
 &&  + \sqrt{(-4 \alpha _1 \lambda_+ ^2-2 \alpha _2 \lambda_+ +\lambda_+ ^2+6)^2-8 (-3 \alpha _1 \lambda_+ ^4-2 \alpha _2 \lambda_+ ^3-18 \alpha _1 \lambda_+ ^2-\alpha _3 \lambda_+ ^2-12 \alpha _2 \lambda_+ -6 \alpha _3)}]
  \end{eqnarray*}

 \end{widetext}
\end{document}